\def\ba{\begin{eqnarray}}
\def\ea{\end{eqnarray}}
\def\l{\label}
\def\n{\nonumber \\}
\def\b{\bibitem}

\def\p{\bar{p}}
\def\erf{\rm erf}

\documentclass[12pt]{article}
\usepackage{epsfig}
 \begin{document} 
\title{Balance of baryon number\\ in the quark coalescence model}

\author{A.Bialas \\ M.Smoluchowski Institute of Physics\\ Jagellonian
University, Cracow\thanks{Address: Reymonta 4, 30-059 Krakow, Poland;
e-mail:bialas@th.if.uj.edu.pl;} \\[0.2cm] and\\[0.2cm] J.Rafelski\\ 
Department of Physics, University of Arizona\thanks{1118 E. 4th Street, Tucson, AZ  85721, USA;
e-mail:rafelski@physics.arizona.edu}}
\date{August 5, 2005}
\maketitle

\begin{abstract}

The charge and baryon balance functions are studied in the coalescence
hadronization mechanism of quark-gluon plasma. Assuming that in the
plasma phase the $q\bar{q}$ pairs form uncorrelated clusters whose decay
is also uncorrelated, one can understand the observed small width of the
charge balance function in the Gaussian approximation. The coalescence
model predicts even smaller width of the baryon-antibaryon balance
function: $\sigma_{B\bar{B}}/\sigma_{+-}= \sqrt{2/3}$.

\end{abstract}

\section*{\normalsize 1. Introduction}\label{sec1} It was recently
argued \cite{bialbal} that the constituent quark coalescence model
\cite{bud} can explain the small pseudo-rapidity width of the charge
balance function \cite{isr,dp} observed in central collisions of heavy
ions at RHIC \cite{star}, as compared to that measured in pp collisions
\cite{isr}. The charge (or baryon number) balance function measures the
relative distance between produced positive and negative charges.
Encouraged by the success of this simple argument, in the present paper
we apply the same idea to the balance function measuring the
baryon-antibaryon compensation.

Our aim is to discuss correlations between produced hadrons, and thus
the quark coalescence model must be generalized to consider possible
preexistent correlations inside the system. In a state rich in
quark-antiquark pairs arising from initial (charge and flavor neutral)
gluon fragmentation we can supplement the standard formulation of the
model by the hypothesis that -before hadronization- the QGP forms
charge- and flavor-neutral $q\bar q$-clusters, which seems natural in a
gluon-dominated plasma. In the final, pre-hadronization stage the
clusters dissociate, quarks and antiquarks coalesce into observed hadrons
while gluons required to balance color charge form additional neutral
quark clusters (and again gluons) and the process is repeated till
hadronization is completed. 

For this reason we can assume for the purpose of evaluating the balance
correlation that at the last stage before hadronization the plasma
consists of isotropic $q\bar{q}$ clusters which are charge-, flavor- and
(though natural, here important) baryon number- neutral. In the
coalescence model a charged meson is formed from two such $q\bar{q}$
clusters. The charged bosons final state balance function is thus
reduced by a factor $1/\sqrt{2}$ with respect to the width of source
distributions, since as is well known, the width of the distribution of
the average of any two independent variables ($q$, $\bar q$ in different
$q\bar{q}$) is reduced by the factor $1/\sqrt{2}$ with respect to the
width of their own distributions. This was the argument in
Ref.\,\cite{bialbal}.

Here we observe that since baryons are formed from three constituent
quarks, we expect in this case a reduction by a factor $1/ \sqrt{3}$,
and thus an even further reduction (by the factor $ \sqrt{2/3}$ beyond
charge balance narrowing) of the width of the baryon balance function. A
particularly interesting case is the balance function of neutral strange
baryons ($\Lambda(sud)$) since the coalescence formation requires here
under any possible microscopic mechanism the participation of three
independent ($s\bar s,\,u\bar u,\, d\bar d$) neutral quark clusters.

The confirmation of this prediction by the future data could be
considered as an important indication that the coalescence model is the
dominant mechanism in the hadronization of the quark-gluon plasma. The
coalescence model is the key ingredient required to describe the
relative yields of hadrons within the statistical hadronization
approach~\cite{share}: the probability of production of a hadron is by
assumption a product of probabilities of finding the constituent quarks
of the hadron considered. 

With this formulation all consequences of the coalescence model
concerning single particle distributions are of course unchanged. In
particular, the spectacular successes of the model in description of
single particle spectra \cite{bud,bp} and flow parameters \cite{wol} in
the central collisions of heavy ions remain valid. Thus our extension is
a natural attempt to make the coalescence model more complete.

To proceed, we need to define some properties of the neutral clusters.
To keep the discussion as simple as possible, we shall assume that (i) a
cluster decays isotropically in its rest frame, and (ii) momenta of the
decay products are uncorrelated. These two properties are sufficient to
derive most of our conclusions. To keep the problem as simple as
possible, we shall also assume that the clusters themselves are
uncorrelated, as would be expected if these arise from gluons.

Using this scenario, we give predictions for the (pseudo)rapidity width
of the charge and baryon balance functions, including corrections due to
the finite acceptance region. In the next section \ref{sec2} we
introduce the notation. The analysis of the charge balance function
given in \cite{bialbal} is shortly summarized in section 3. Estimate of
the baryon balance function is given in section 4. Our conclusions are
listed in the last section.

\section*{\normalsize 2. Balance functions}\label{sec2}

In this paper we shall only discuss the neutral systems, u.e. the
systems with vanishing  total charge and baryon number. In this case
the balance function of the ``charges'' denoted by $+$ and $-$ is
 expressed in terms of the single- $N$ and double- $D$ 
particle functions \cite{dp,jp} as follows:
\ba
B(\Delta_2|\Delta_1)&=&
\frac{D(+,\Delta_2|-,\Delta_1)-D(+,\Delta_2|+,\Delta_1)}{N_+(\Delta_1)}
+\n[0.2cm]
&+&
\frac{D(-,\Delta_2|+,\Delta_1)-D(-,\Delta_2|-,\Delta_1)}{N_-(\Delta_1)} .
\l{9} 
\ea 
Here
\ba
D(i,\Delta_2|j,\Delta_1)=
\int_{\Delta_2} dy_2 \int_{\Delta_1} dy_1
\frac{d^2n_{ij}}{dy_{1}dy_{2}}    \l{7}
\ea
\ba 
N_i(\Delta)=\int_{\Delta} \frac {dN_i}{dy} dy  \l{7a}
\ea
where  $dn_i/dy$ and $dn^2_{ij}/dy_1dy_2$ are the corresponding particle densities.

The measurement of the electric charge balance function performed by the STAR 
collaboration requires both particles to be in a finite rapidity acceptance
interval $-\Delta\leq y_1,y_2\leq \Delta$   while the difference $\delta$ of
(pseudo)rapidities is kept {\it fixed}. This suggests a change of
variables:
\ba
y_1-y_2 \equiv \delta; \qquad  {y_1+y_2\over 2} \equiv z. \l{ix}
\ea

With this notation, the integration in Eq.\,(\ref{7}) must be restricted 
to $\delta$ being fixed, while $z$ must be kept
inside the interval $-\bar{\Delta}\leq z\leq \bar{\Delta}$ where
\ba
\bar{\Delta}= \Delta-{|\delta|\over 2}.   \l{ixa}
\ea 
We shall use this prescription also for the baryon number 
which is of interest in this paper.

Thus the balance function we discuss is defined as
\ba
B_s(\delta;\Delta)&=& \frac{\int_{-\bar{\Delta}}^ {-\bar{\Delta}}dz \left[
d^2n_{+-}/dzd\delta -d^2n_{++}/dzd\delta\right]}
{\int_{-\Delta}^{+\Delta} dy_+ dn_+/dy_+} + \n[0.2cm]
&+&
\frac{\int_{-\bar{\Delta}}^ {-\bar{\Delta}}dz \left[
d^2n_{-+}/dzd\delta -d^2n_{--}/dzd\delta\right]}
{\int_{-\Delta}^{+\Delta} dy_- dn_-/dy_-}  \l{i5}
\ea
where $\pm$ refers here to the baryon number.

\section*{\normalsize 3. Balance function in the coalescence model}
\label{sec3}
 We first consider the production of charged pions
\cite{bialbal}. In formation of such a pair, the dominant process is
that involving one $U$-cluster and one $D$-cluster:
\ba
(u\bar{u})+ (d\bar{d})\; \rightarrow\; (u\bar{d})+ (\bar{u}d).   \l{m1}
\ea
The corresponding two particle momentum distribution is:
\ba
\rho(p_+,p_-)&=&\hspace*{-0.2cm}\int dP_1dP_2 \rho_c(P_1)\rho_c(P_2)
 \int dp_{u_1}d\p_{u_1}dp_{d_2}d\p_{d_2}\n
\hspace*{-0.6cm} &&f(P_1,p_{u_1})f(P_1,\p_{u_1})f(P_2,p_{d_2})f(P_2,\p_{d_2})\n
\hspace*{-0.6cm} &&\delta[p_+-(p_{u_1}+\p_{d_2})/2]\delta[p_--(\p_{u_1}+p_{d_2})/2]\n
\hspace*{-0.6cm} &&G_m(p_{u_1}-\p_{d_2})G_m(p_{d_1}-\p_{u_2})   \l{m2}
\ea
where $\rho_c(P)$ is the distribution of clusters, $f(P,p)$
 the cluster dissociation probability, that is
the probability of finding a quark of momentum $p$ in the cluster of momentum $P$,
and $G(p-\p)$ is the probability of coalescence, {\it i.e.} that  a
quark-antiquark pair   forms a hadron. The momenta of quarks are denoted by
$p$, those of antiquarks by $\p$.

Since one pair of clusters cannot produce two mesons of the same charge,
only $\rho(p_+,p_-)$ given by (\ref{m2}) contributes to the balance
function. In the following we shall only consider rapidity and
pseudo-rapidity distributions. To simplify the technicalities we 
take all functions in (\ref{m2}) in the Gaussian form:
\ba
\rho_c(Y)=\frac{N}{A\sqrt{\pi}}e^{-Y^2/A^2}\;;\;\;
f(Y,y)= \frac1{c\sqrt{\pi}}e^{-(Y-y)^2/c^2}\;;\;\;   \l{m3}
\ea
where $N$ is the total number of clusters. It turns out that the 
balance function does not depend on the shape of $G(u)$.

Introducing  (\ref{m3}) into (\ref{m2}) allows to perform all
integrations. The result, substituted  in (\ref{i5}), gives
\ba
B_{m}(\delta;\Delta)=\frac1{c\sqrt{\pi}}e^{-\delta^2/c^2}
\frac{\erf[2\bar{\Delta}/\sqrt{2A^2+c^2}]}
{\erf[2\Delta/\sqrt{2A^2+2c^2}]}   \l{m4}
\ea

One sees that in the limit of large acceptance ($\Delta \rightarrow
\infty$) the width of the balance function is determined by the same
parameter $c$ which describes the cluster dissociation   width. The effect of
finite acceptance, $\Delta$, is seen as the ratio of the two error
functions. Note that in the Gaussian distribution model the balance function
is entirely independent of the coalescence probability distribution $G(u)$.

\section*{\normalsize 4. Baryon balance function}\label{sec4}
The dominant process in the production of baryon-antibaryon
pairs involves three clusters:
\ba
(q\bar{q})+(q\bar{q})+(q\bar{q})\;\rightarrow \; 
(qqq)+(\bar{q}\bar{q}\bar{q})   \l{b1}
\ea
Consequently, the distribution of the $B\bar{B}$ pair is:
\ba
\rho(p,\p)&=&\hspace*{-0.2cm}\int dP_1dP_2dP_3 \rho_c(P_1)\rho_c(P_2)\rho_c(P_3)
\int dp_1dp_2dp_3d\p_1d\p_2d\p_3   \n
\hspace*{-0.6cm} && f(P_1,p_1)f(P_2,p_2)f(P_3,p_3)f(P_1,\p_1)f(P_2,\p_2)f(P_3,\p_3)\n
\hspace*{-0.6cm}&&\delta[p-(p_1+p_2+p_3)/3]\delta[\p-(\p_1+\p_2+\p_3)/3]\n
\hspace*{-0.6cm}&&G_B(p_1-p_2,p_2-p_3,p_3-p_1)G_B(\p_1-\p_2,\p_2-\p_3,\p_3-\p_1).
\qquad  \l{b2}
\ea

Introducing  Eq.\,(\ref{m3}) into Eq.\,(\ref{b2}), performing the integrations and 
putting the result into Eq.\,(\ref{i5}), one obtains:
\ba
B_{B}(\delta;\Delta)=\frac{\sqrt{3}}{c\sqrt{2\pi}}\,e^{-3\delta^2/(2c^2)}\,
\frac{\erf[\sqrt{6}\,\bar{\Delta}/\sqrt{2A^2+c^2}]}
{\erf[\sqrt{3}\Delta/\sqrt{A^2+c^2}]}\,.   \l{b3}
\ea

One sees that in the limit of large acceptance the width of baryon
balance function is again determined by the parameter $c$ but it is by
the factor $\sqrt{2/3}$ smaller than that for obtained for the charged
pion balance functions given by Eq.\,(\ref{m4}). The corrections due to
finite acceptance are also given by ratio of two error functions but
their arguments are multiplied by $\sqrt{3/2}$, as compared to those in
Eq.\,(\ref{m4}). Therefore they are slightly less important than in the pion
case.

 It is natural to expect that the $(q\bar{q})$ cluster dissociation 
probability is isotropic. As discussed in \cite{bialbal}, this puts a strong
constraint on the parameter $c$ responsible for the cluster decay
distribution. Indeed, the isotropic decay implies that the decay
distribution in pseudo-rapidity is 
\ba 
\rho(\eta) \approx \frac1{2(\cosh\eta)^2}\,, \l{i10} 
\ea 
corresponding to the (pseudo)rapidity distribution of statistically produced
(massless) particles. The width of this distribution is
$\langle |\eta|\rangle \equiv \int d\eta |\eta| \rho(\eta)=\log 2\approx 0.69$.
In order to approximate  the distribution (\ref{i10}) by a Gaussian of the 
same width we must use:
\ba 
c =\sqrt{\pi}\log 2 \approx 1.23 \,. \l{n1} 
\ea

Although Eq.\,(\ref{n1}) refers to pseudo-rapidity distribution, we shall take
this value of $c$ to describe as well the rapidity distribution in
cluster decay\footnote{One can show that the parameter $c$ of the
rapidity distribution cannot be larger that that describing the
pseudo-rapidity distribution.}. As shown in \cite{bialbal}, this value is
compatible with the STAR data \cite{star}, provided that the finite
acceptance corrections are included ($A=3.5$ was assumed). These
corrections are important: they reduce the width from $0.69$ to $0.55$,
compatible with data.

\begin{figure}[htb]
\centerline{%
\epsfig{file=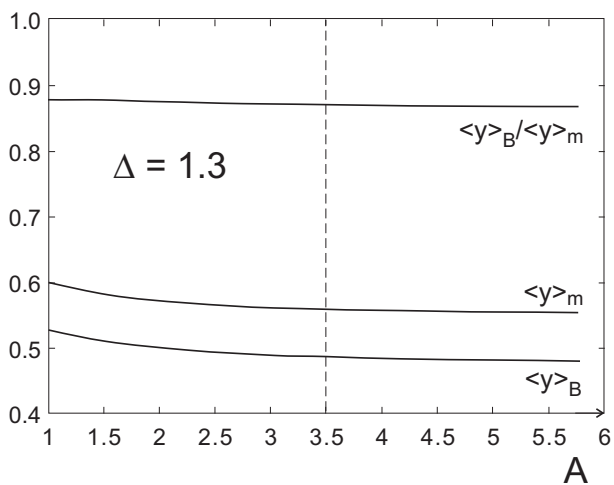,width=8cm}}
\caption{Widths of  balance functions versus the width  $A$
of the pre-hadronization  cluster rapidity distribution.
\label{fig1}}
\end{figure}

The comparison of the results for charge and baryon balance functions is
shown in Fig. \ref{fig1} where the expected widths, calculated from Eq.\,(\ref{m4})
and Eq.\,(\ref{b3}) with $c$ given by Eq.\,(\ref{n1}) and $\Delta = 1.3$, are
plotted as function of the cluster distribution width $A$, to show (in)sensitivity 
with regard to the cluster distribution.  Also in Fig. \ref{fig1}  
the ratio of the charge (m: meson) and baryon balance function widths is shown.

As discussed in Ref.\,\cite{bialbal},  
$\langle |y|\rangle $ for pions agrees reasonably well with the STAR
data in a rather broad range of $A$. One also sees that the ratio of
baryon and pion widths is predicted to be $.87$ and is insensitive to
$A$. This value is larger than $\sqrt(2/3)\approx .82$, expected for
large rapidity acceptance. 

\begin{figure}[htb]
\centerline{%
\epsfig{file=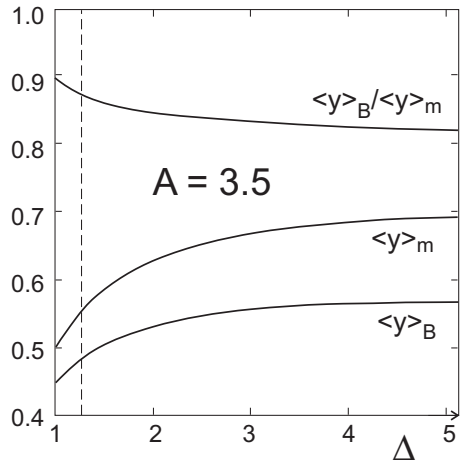,width=6cm}}
\caption{Widths of  balance functions versus the  acceptance
$\Delta$. Vertical dashed line indicates the STAR acceptance.
\label{fig2}}
\end{figure}

In Fig. \ref{fig2} the same quantities are plotted versus  $\Delta$,
the size of the acceptance region for $A=3.5$. One sees that the widths
change rather rapidly when $\Delta$ varies from 1 up to 3. Starting from
$\Delta =4$ the effect of acceptance practically disappears. 

\section*{\normalsize 5. Comments and conclusions}\label{sec5}
Several comments are in order. 

(i) We have ignored the transverse motion of the clusters (the
distribution (\ref{i10}) is only valid for clusters with vanishing
transverse momentum). The effect of transverse motion is expected to
reduce the decay width in rapidity~\cite{z}. This effect was discussed in
Ref.\,\cite{bialbal} and shown to be small for transverse velocities up to 0.4\,.
It may be interesting, however, to investigate -both theoretically and
experimentally- the balance functions at larger transverse momenta. Such
studies may provide additional information about the properties of the
transverse flow.

(ii) We have also ignored the difference between rapidity and
pseudo-rapidity. Therefore, our predictions of the balance function width
in rapidity could be somewhat overestimated. The width ratio between baryons and
pions is, however, much less sensitive to this effect and thus remains a 
much more solid prediction.

(iii) The correlations in cluster decay were neglected. As shown in
\cite{bialbal}, introducing correlations may change somewhat the width
of the balance functions. This may be  an interesting point for future
research.

(iv) As we were using  Gaussian approximation, we cannot discuss
seriously the shape of the balance function. Although the present data
are consistent with Gaussian shapes, it would be interesting to investigate
the problem more thoroughly when better data become available.

(v) We have only considered the dominant processes leading to meson and
baryon formation. The corrections from sub-dominant ones give, generally,
contributions to balance functions which are much broader than those
described in the present paper. 

(vi) The effect of resonance decay was entirely neglected. This serious
problem seems, however, beyond reach of our approach. 
In this context, it should be emphasized that the
argument presented in this paper applies to any kind of
baryon-antibaryon pairs. Thus we feel that measurements of (multi)strange
baryons may be particularly interesting, because -due to their large
mass- they are less sensitive to corrections related to resonance decays
and transverse motion.

In conclusion, we have calculated the width of the charge and baryon
number balance functions in the coalescence model generalized to include
possibility of correlations between constituent quarks and antiquarks in
the form of charge- and flavor- neutral, isotropically dissociating
clusters. The existing data on charge balance functions are explained.
The width of the baryon balance function at large acceptance is
predicted to be smaller by nearly a factor $\sqrt{2/3}$. Finite acceptance
corrections were estimated. We believe  that future measurements of the
baryon balance functions may provide an interesting test of the quark
coalescence hadronization model.

\section*{\normalsize Acknowledgments} Discussions with  M.Lisa and K.Zalewski 
are appreciated. Work supported in part by a grant from: 
the U.S. Department of Energy  DE-FG02-04ER4131.

\end{document}